\documentclass[aps,prc,twocolumn,superscriptaddress,showpacs,showkeys]{revtex4}
\usepackage{graphicx}

\begin{document}


   \title{{\em Ab-initio} shell model with a core}




\author{A. F. Lisetskiy}
\email[]{lisetsky@physics.arizona.edu}
\author{B. R. Barrett}
\author{M.K.G. Kruse}
\affiliation{Department of Physics, University of Arizona, Tucson, AZ 85721}

\author{P. Navratil}
\affiliation{ Lawrence Livermore National Laboratory,
Livermore, CA 94551}
\author{I. Stetcu}
\affiliation{Los Alamos National Laboratory, Los Alamos, NM 87545}
\author{J. P. Vary}
\affiliation{Department of Physics and Astronomy, Iowa State University, Ames,
Iowa 50011}


\date{\today}

\begin{abstract}
We construct effective 2- and 3-body Hamiltonians for the p-shell
by performing $12\hbar\Omega$ {\em ab initio} no-core shell model (NCSM)
calculations for A=6 and 7 nuclei and explicitly projecting the
many-body Hamiltonians onto the $0\hbar\Omega$ space.  We then
separate these effective Hamiltonians into 0-, 1- and 2-body
contributions (also 3-body for A=7) and analyze the systematic
behavior of these different parts as a function of the mass
number A and size of the NCSM basis space. The role of effective
3- and higher-body interactions for $A>6$ is investigated and
discussed.
\end{abstract}


\maketitle

\section{Introduction}
 Microscopic {\it ab-initio} many-body approaches have significantly progressed in recent years
\cite{Nav07,Nog06,Ste05,Nav03,Nav00,Pie02,Pie01,Kow04}. Nowdays, due to
 increased computing power and novel techniques, {\it ab-initio} calculations  are able to reproduce
 a large number of observables for atomic nuclei with  mass up to A=14. The light nuclei have also served
 as a crucial site to recognize the important role of three-body forces and three-body correlations.
Approaches like the No-Core Shell Model (NCSM) \cite{Nav00}, the Green's Function Monte Carlo (GFMC) \cite{Pie02}
and  the Coupled-cluster theory with single and double excitations (CCSD) \cite{Kow04} can be formally extended for
heavier nuclei. However, the explosive growth in computational power, required to achieve convergent
results, severely hinders the detailed {\it ab-initio} studies of heavier, A$\ge16$, nuclei. In the case of the NCSM, the slow
convergence of the calculated energies is caused by the adoption of a two-body cluster approximation,
which does not take many-body correlations into account. Straightforward employment of the three-body and
higher-body interactions dramatically complicates the problem, even for light nuclei.

An alternative approach is to construct a small-space effective two-body interaction, which would account for
the many-body correlations for the A-body system in a large space. Attempts to include many-body correlations
approximately modifying the one-body part of the effective two-body Hamiltonian and
 employing  a unitary transformation have been reported recently \cite{Fuj07}.

In this paper we derive a valence space ($0\hbar\Omega$) effective two-body interaction that accounts
for all the core-polarization effects available in the {\em ab-initio} NCSM wavefunctions.

First, in the framework of the NCSM, we construct the effective Hamiltonians on the two-body cluster level
for A=6 systems in the $N_{\rm max}\hbar \Omega$ space. $N_{\rm max}$ represents the limit on the total oscillator
quanta (N) above the minimum configuration. We take  $N_{\rm max}$ values from 2 to 12.
Second, following the original idea of Ref. \cite{Nav97},  we employ an unitary many-body transformation
and obtain the effective two-body Hamiltonian in the $0\hbar \Omega$ space (p-shell space), which exactly reproduces the lowest,
$0\hbar \Omega$ space dominated, eigenstates of the 6-body Hamiltonian  in the large $N_{\rm max}\hbar \Omega$
space. Third, we perform NCSM calculations for A=4 and A=5 systems with the effective Hamiltonian constructed
on the two-body cluster level for the A=6 system and determine the core and one-body parts of the
effective two-body Hamiltonian for A=6 in the p-shell space. Finally, the procedure is generalized for arbitrary
mass number A.  We analyze the properties of the constructed two-body Hamiltonians, investigate
their efficiency to reproduce the observables of different A-body systems calculated in large
$N_{\rm max}\hbar \Omega$ spaces and study the role of the effective p-shell space three-body interaction.

\section{Approach}
 \subsection{No Core Shell Model and effective interaction}
 The starting point of the No Core Shell Model (NCSM) approach is the bare,
 exact  A-body Hamiltonian  constrained by the Harmonic Oscillator (HO)
 potential \cite{Nav00}:

 \begin{equation}
 \label{hOmA}
 H^{\Omega}_{A} = \sum_{j=1}^{A}h^{\Omega}_j + \sum_{j>i=1}^{A}V_{ij}(\Omega,A),
 \end{equation}
  where $h^{\Omega}_j$ is the one-body HO Hamiltonian
  \begin{equation}
 h^{\Omega}_j = \frac{p^2_j}{2m} +\frac{1}{2}m\Omega^2r^2_j
 \end{equation}
 and $V_{ij}(\Omega,A)$ is a bare NN interaction $V^{\rm NN}_{ij}$ modified
 by the term introducing A- and $\Omega$-dependent corrections  to offset the HO potential present
in $h^{\Omega}_j$:
  \begin{equation}
  \label{vOmA}
  V_{ij}(\Omega,A)=V^{NN}_{ij}-\frac{m\Omega^2}{2A}(\vec{r}_i-\vec{r}_j)^2.
 \end{equation}
 The eigenvalue problem for the exact A-body Hamiltonian (\ref{hOmA}) for $A>3$ is
 very complicated technically, since an extremely large A-body HO basis is required to
 obtain converged results. However, the $A=2$ problem  is
 considerably simpler. For many realistic NN interactions its solution in the relative HO basis with $N_{\rm max}=450$
 accounts well for the short range correlations and is a precise approximation for
 the infinite space ($N_{\rm max} \rightarrow \infty $) result. This allows one to adopt the two-body
 cluster approximation to construct the NCSM effective two-body Hamiltonian
 $H^{N_{\rm max}, \Omega}_{A,a=2}$ for an A-body system in an $N_{\rm max} \hbar \Omega$ space
  of tractable dimension, where the lower index $a$ stands for the number of particles in the cluster.
  This approximation consists of solving Eq.(\ref{hOmA}) for the $a=2$ body subsystem of A leading to
  \begin{equation}
   \label{hOmA2}
 H^{\Omega}_{A,a=2} = h^{\Omega}_1+h^{\Omega}_2 + V_{12}(\Omega,A).
 \end{equation}
  The information about the total number of interacting particles A
 enters the  bare $H^{\Omega}_{A,a=2}$ Hamiltonian through the second term in the right hand side of
(\ref{vOmA}).
Next, we find the unitary transformation $U_2$ which reduces the bare  $H^{\Omega}_{A,a=2}$ Hamiltonian in
the ``infinite space'' ($N_{\rm max}^{\infty}=450$)
 to the diagonal form:
 \begin{equation}
   \label{hOmA2e}
 E^{\Omega}_{A,2} = U_2 H^{\Omega}_{A,2} U_2^\dagger,
 \end{equation}
where, for the sake of simplicity, we omit the index $A$ for $U_2$ and keep only the index $a$ indicating the
order of cluster approximation.
  The matrix $U_2$ can be split into 4 blocks:
 \begin{eqnarray}
 \label{udec}
U_2 =
\left(\begin{array}{cc}
U_{2,P} & U_{2,PQ} \\
U_{2,QP} & U_{2,Q} \\
\end{array} \right),
\end{eqnarray}
 where the square $d_P\times d_P$ $U_{2,P}$ matrix corresponds to the P-space (or model space) of dimension $d_P$,
characterized by the chosen $N_{\rm max}$ value.

 Taking into account that the $E^{\Omega}_{A,2}$ matrix has a diagonal form
 \begin{eqnarray}
 \label{edec}
E^{\Omega}_{A,2} =
\left(\begin{array}{cc}
E^{\Omega}_{A,2,P} & 0 \\
0 &  E^{\Omega}_{A,2,Q}\\
\end{array} \right),
\end{eqnarray}
  one can calculate the effective  $H^{N_{\rm max},\Omega}_{A,2}$ Hamiltonian using the
 following formula:

\begin{equation}
 \label{heff1}
 H^{N_{\rm max},\Omega}_{A,2} =
              {U_{2,P}^\dagger \over \sqrt{U_{2,P}^\dagger U_{2,P}}}
              E^{\Omega}_{A,2,P}
              {U_{2,P} \over \sqrt{U_{2,P}^\dagger U_{2,P}}}.
\end{equation}
It is easy to show by inserting  Eq.(\ref{hOmA2e}) into the Eq.(\ref{heff1}), and taking into account
 Eq.(\ref{udec}) that the unitary transformation (\ref{heff1}) is equivalent to the commonly used
unitary transformation \cite{Oku54,Suz82} and that Eq.(\ref{heff1}) is identical to the Eqs.(15,16) from \cite{Nav00}.
We note, that, by using Eq.(\ref{heff1}) one does not need to calculate and store a large number of matrix elements of
 the $\omega$-operator (i.e., $U_{2,P}^\dagger \omega_2 = U_{2,PQ}^\dagger$). Furthermore, the decoupling condition
$QH^{\rm eff}P=0$ is automatically satisfied, which is obvious from the diagonal form of
the $E^{\Omega}_{A,2}$ matrix. We note that our treatment of center-of-mass motion remains the same as in the NCSM
(Ref. \cite{Nav00}). We initiate all effective interaction developments at the A-body level, and,
through a series of steps, arrive at a smaller space effective interaction appropriate for the A-body system.
 For this reason, our derived effective Hamiltonians have their first subscript as ``A''.

\subsection{\label{proj} Projection of the many-body Hamiltonian}

The next step of the traditional NCSM prescription is to construct the full A-body Hamiltonian using
the effective two-body Hamiltonian (\ref{heff1}) and to diagonalize it in the A-body $N_{\rm max}$ model
space. As we increase the number of nucleons, the dimension of the corresponding
$N_{\rm max}$ model space increases very rapidly. For instance, up-to-date computing
resources allow us to go as high as $N_{\rm max}=16$ for the lower part of the p-shell (A=5,6) \cite{A56},
 while already for the upper part of the p-shell (A$\sim$15), we are limited to  $N_{\rm max}=8$.
The computational eigenvalue problem for many-body systems is complicated  because of the very large
matrix dimensions involved. The largest dimension of the model space that we encountered in this study
for $^6$Li with $N_{\rm max}=12$ exceeds $d_P=4.8\times10^7$. To solve this problem we have used the
specialized version of the shell-model code ANTOINE \cite{Cou99,Mar99}, recently adapted for the NCSM
\cite{Cou01}.

In fact, the NCSM calculation for the A=6 system in the $N_{\rm max}=12$ space yields nearly converged
energies for the lowest states dominated by the $N=0$ components, while there is incomplete
convergence for $A \ge 15$ in $N_{\rm max}=8$ space. Therefore, considering the $N_{\rm max}=12$ NCSM
results as exact solutions for the lowest $N=0$ dominated 6-body states, we may construct the
$N_{\rm max}=0$ space Hamiltonian for the A=6 system, which exactly reproduces those  $N_{\rm max}=12$
 eigenvalues \cite{Nav97}.
Moreover, if it is possible to solve the 6-body problem for A=6, then it is possible to solve the 6-body problem
for arbitrary A, using the corresponding effective Hamiltonian $H^{N_{\rm max},\Omega}_{A,2}$ obtained in
the two-body cluster approximation. This means that we can determine for any A-body system the effective
Hamiltonian in the $N_{\rm max}=0$ space, which accounts for 6-body cluster
dynamics in the large $N_{\rm max}=12$ space.

To generalize, we start by defining the procedure for determining the effective Hamiltonian matrix elements for the $a_1$-body cluster
in the A-nucleon system. We do this by constructing the full $a_1$-body Hamiltonian using the effective 2-body Hamiltonian
(\ref{heff1}) and diagonalizing it in the  $N_{\rm max}$ model space. In the spirit of Eq.(\ref{hOmA2}), this yields the
eigenenergies  $E^{N_{\rm max},\Omega}_{A,a_1}$ of the $a_1$-body system and their corresponding  $a_1$ eigenvectors which make up the unitary
transformation matrix  $U_{a_1,P}^{A,N_{\rm max}}$. These $a_1$-body results can then be projected into a smaller,
secondary $P_1$-space,  given by $N_{1,\rm max} \hbar \Omega$ with $N_{1,\rm max}=0$, where, similar to Eqs.(\ref{udec})
 and (\ref{edec}),   $E^{N_{\rm max},\Omega}_{A,a_1}$
and  $U_{a_1,P}^{A,N_{\rm max}}$ can be split into parts related to the two spaces, $P_1$ and $Q_1$,
 where $P_1+Q_1=P$. The new secondary effective Hamiltonian then takes the following general form:

\begin{equation}
 \label{heff2}
 {\cal H}^{N_{1,\rm max}, N_{\rm max}}_{A,a_1} =
              {U_{a_1,P_1}^{A,\dagger} \over \sqrt{U_{a_1,P_1}^{A,\dagger} U_{a_1,P_1}^{A}}}
              E^{N_{\rm max},\Omega}_{A,a_1,P_1}
              {U_{a_1,P_1}^{A} \over \sqrt{U_{a_1,P_1}^{A,\dagger} U_{a_1,P_1}^{A}}},
\end{equation}
where the $\Omega$ superscript on the left-hand side is omitted for the sake of simplicity.
As stated earlier, the new index $a_1$ determines the order of the cluster approximation  in
the smaller $P_1$ space, {\em i.e.}, $N_{1,\rm max}=0$. Because the $P_1$ space has $N_{1,\rm max}=0$,
the projection into this space "freezes" some number
of the $a_1$ nucleons into fixed single particle configurations, which can be thought of as the "inert core" states in
the Standard Shell Model (SSM) approach. Consequently, it is possible to write $a_1$ as
$a_1 = A_c + a_{\rm v}$,  where $A_c$ is the number of nucleons making up the core configuration, while $a_{\rm v}$ refers
to the size of valence cluster.

 For instance, in the case of p-shell nuclei, $A_c=4$, so, if $a_1=5$ ({\em i.e.} the 5-body cluster approximation), then
 the effective Hamiltonian ${\cal H}^{N_{1,\rm max}=0, N_{\rm max}}_{A,a_1=5}$ is simply a one-body Hamiltonian ($a_{\rm v}=1$)
 appropriate for the A-nucleon system. Similarly, for the 6-body cluster approximation, {\em i.e.}, $a_1=6$, we obtain the
 effective Hamiltonian ${\cal H}^{N_{1,\rm max}=0, N_{\rm max}}_{A,a_1=6}$, which is a two-body Hamiltonian
 ($a_{\rm v}=2$)
 for the A-body system, and, so on for larger values of $a_1$.  Whatever the value of $a_{\rm v}$ is, the effective Hamiltonian
 ${\cal H}^{N_{1,\rm max}=0, N_{\rm max}}_{A,a_1}$ contains the information about the $a_1$-body dynamics in the original
large $N_{\rm max} \hbar \Omega$ space, since it reproduces exactly the  lowest $d_{P_1}$ eigenvalues $E^{N_{\rm max},\Omega}_{A,a_1,P_1}$
of the $a_1$-body Hamiltonian in the $N_{\rm max} \hbar \Omega$ space,
 where  $d_{P_1}$ is a dimension of the $P_1$ space.

In the case of a doubly magic closed shell with two extra nucleons i.e., $A=6,18,42$, {\em etc.},
the dimension of the effective Hamiltonian ${\cal H}^{0,N_{\rm max}}_{A,a_1=A}$ is a 2-body ($a_{\rm v}=2$)
 Hamiltonian in the p-, sd-, pf-spaces, {\em etc.}, respectively. This means that the secondary effective
 Hamiltonian (\ref{heff2}) contains only 1-body and 2-body terms, even after the {\it exact} A-body cluster
transformation. This effective Hamiltonian (\ref{heff2}), which now contains the correlation energy of \underline{all}
 A nucleons, is the correct one-body plus two-body Hamiltonian to use in a SSM calculation with inert core.
The $A_c=A-2$ nucleon-spectators fully occupy the shells below the valence shell and the total A-body wave-function can
 be exactly factorized as the $A_c$-body "core" and the valence 2-body wave functions.
 This considerably simplifies the calculation of the effective Hamiltonian, because only the
$0\hbar \Omega$ part (P$_1$-space part) of the complete $N_{\rm max}\hbar \Omega$ wave function needs to be specified.


\section{Effective two-body p-shell interaction}

\begin{figure}[!ht]
 \includegraphics[scale = 0.21]{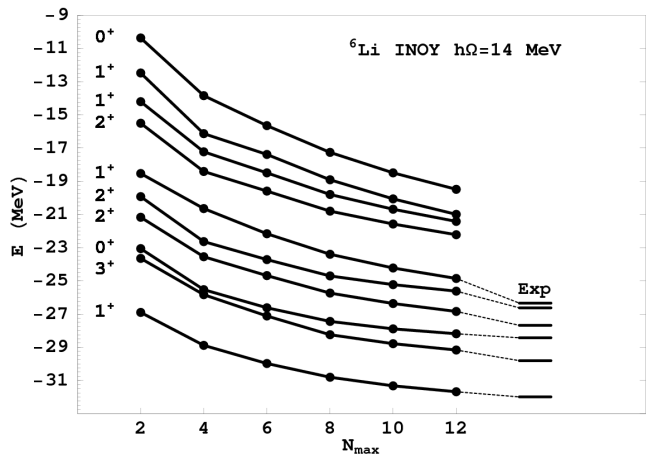}%
 \caption{\label{spectra6li14hw} The excitation energies of the $J^\pi$ states and ground state energy
for $^{6}$Li calculated in  the $N_{\rm max}\hbar \Omega$ spaces with the INOY interaction and $\hbar \Omega=14$ MeV.
 The experimental spectra and ground state energy are shown for comparison.}
 \end{figure}

 Utilizing the approach outlined above, we have calculated effective p-shell Hamiltonians for $^{6}$Li,
using the 6-body Hamiltonians with $N_{\rm max}=2,4,..,12$ and $\Omega=14$ MeV, constructed from the
INOY (inside nonlocal outside Yukawa) interaction \cite{Dol04,Dol03}.  This is a new type of interaction,
which has local behavior appropriate for traditional NN interactions at longer ranges, but exhibits a
nonlocality at shorter distances. The nonlocality of the NN interaction has been introduced in order
to account effectively for three-nucleon (NNN) interactions which correctly describe the NNN bound
states $^3$H and $^3$He, whereas still reproducing NN scattering data with high precision.
The corresponding excitation energies of p-shell dominated states and
the binding energy of $^{6}$Li are shown in Fig.\ref{spectra6li14hw} as a function of $N_{\rm max}$.
The dimension of the configurational space for the $N_{\rm max}=12$ case considered is 48 million (M-scheme).
A two orders of magnitude increase in the size of the model space, as compared to the previous
$N_{\rm max}=6$ study \cite{Nav97},  allows us to determine a converged value of 31.681 MeV
for the $^6$Li binding energy. Furthermore, the excitation energy of the highest lying p-space state,
$J_\pi=0^+_2$, is lowered by an amount of 2.1 MeV in comparison to the $N_{\rm max}=6$ case, indicating
improved convergence for both the excited states and ground state for $N_{\rm max}=12$.

In the SSM an effective two-body Hamiltonian for a nucleus with mass number A is represented
in terms of three components:
\begin{equation}
\label{hssm}
H_{\rm SSM}^A= H_0 + H_1 + V_2^A,
\end{equation}
where $H_0$ is the inert core part associated with the interaction of the nucleons occupying closed shells,
$H_1$ is the one-body part corresponding to the interaction of valence nucleons with core nucleons, and
$V_2^A$ is the two-body part referring to the interaction between valence particles.
  It is usually assumed that  the core and one-body parts are constant for an arbitrary number of valence particles
 and that only the two-body part $V_2^A$ may contain mass dependence that includes effects of three-body and higher-body interactions.

To represent the ${\cal H}^{0,N_{\rm max}}_{A,a_1}$ Hamiltonian in the SSM format,
we develop a valence cluster expansion (VCE),
\begin{equation}
\label{hexp}
{\cal H}^{0,N_{\rm max}}_{A,a_1}=H_0^{A,A_c}+H_1^{A,A_c+1}+\sum_{k=2}^{a_{\rm v}}V_k^{A,A_c+k},
\end{equation}
where the lower index, k, stands for the k-body interaction in the $a_{\rm v}$-body valence cluster ($a_1=A_c+a_{\rm v}$);
 the first upper index A for the mass dependence; and the second upper index, $A_c+k$ for the number of particles
 contributing to the corresponding k-body part. Thus, we consider the more general case of allowing the core (k=0),
 one-body (k=1) and other k-body parts to vary with the mass number A. This appears necessary to include the A-dependence
 of the excitations of the core ($A_c$) nucleons treated in the original N$_{\rm max}$ basis space.
  For the A=6 case the two-body valence cluster (2BVC) approximation is exact:
 \begin{equation}
\label{hncsm}
{\cal H}^{0,N_{\rm max}}_{A=6,a_1=6}= H_{0}^{6,4} + H_{1}^{6,5} + V_{2}^{6,6},
\end{equation}
where the core part, $H_{0}^{6,4}$, is defined as the ground state $J^\pi=0^+$  energy of $^4$He calculated in
the $N_{\rm max} \hbar \Omega$ space with the TBMEs of the primary effective Hamiltonian,
$H^{N_{\rm max},\Omega}_{6,2}$ for A=6.
Then the one-body part, $H_{1}^{6,5}$,is determined as
 \begin{equation}
\label{v1}
 H_{1}^{6,5}={\cal H}^{0,N_{\rm max}}_{6,5}-H_{0}^{6,4}.
\end{equation}
The TBMEs of the one-body part, $H_{1}^{6,5}$,
\begin{equation}
\label{v1me}
 \langle ab| H_1^{6,5}| cd \rangle_{JT} = ( \epsilon_a + \epsilon_b ) \delta_{a,c} \delta_{b,d}
\end{equation}
 may be represented in terms of single particle energies (SPE) , $\epsilon_a$:
\begin{equation}
\label{v1spe}
\epsilon_a^p=  E(^5\mbox{Li},j_a) - H_0^{6,4}, \\
\epsilon_a^n=  E(^5\mbox{He},j_a) - H_0^{6,4}.
\end{equation}
where the index a (as well as b,c, and d) denotes the set of single particle HO quantum numbers $(n_a,l_a,j_a)$, 
upper index stands for proton (p) and neutron (n), and
the E($^5$Li,J), E($^5$He,J)  are NCSM energies of the lowest $J^\pi_i=3/2^-_1$ and $J^\pi_i=1/2^-_1$ states calculated
in the $N_{\rm max}\hbar \Omega$ space for the $5$-body system using the TBMEs of the $A=6$ effective
Hamiltonian, $H^{N_{\rm max},\Omega}_{A=6,2}$, which includes Coulomb energy.
Finally, the two-body part $V_2^{6,4}$ is obtained by subtracting of two Hamiltonians:
 \begin{equation}
\label{v2}
 V_{2}^{6,6}={\cal H}^{0,N_{\rm max}}_{6,6}-{\cal H}^{0,N_{\rm max}}_{6,5}.
\end{equation}
 It is worth noting that since the Coulomb energy is included in the original Hamiltonian, the proton-proton (pp),
neutron-neutron (nn) and proton-neutron (pn) $T=1$ TBMEs of the two-body part, $V_{2}^{6,6}$, have different values.
The pn TBMEs of the core, one-body and two-body parts of the expanded Hamiltonian for $^6$Li are listed in the
Table I.
 \begin{table*}[tbp]
 \caption{\label{V6dec} The pn TBMEs of the NCSM  $H^{N_{\rm max}=12, \Omega}_{A=6,2}$ Hamiltonian with $\Omega =14$ MeV, the p-shell effective Hamiltonians ${\cal H}^{0,N_{\rm max}}_{6,6}$
and ${\cal H}^{0,N_{\rm max}}_{7,6}$ obtained from an $N_{\rm max}=12$ NCSM calculation for $^6$Li are shown.
The core, $H_{0}^{A,4}$, one-body, $H_1^{A,5}$,  and residual two-body,$V_2^{A,6}$ parts for latest two
Hamiltonians are presented. The ${\cal H}^{0,N_{\rm max}}_{7,6}$ Hamiltonian with A-independent core and
one-body parts is shown in last three columns.}
 \begin{tabular}{cc|cc|cc|rrr|rrr|rrr|rrr}
 \hline\hline
&&&&&&&&&&&&&&&&& \\
      &      &      &       &   &   &$H^{12,\Omega}_{A,2}$
                                     &\multicolumn{2}{ c|}{${\cal H}^{0,12}_{A,6}$,  (MeV)}  &
                                       \multicolumn{3}{|c|}{${\cal H}^{0,12}_{6,6}$, (MeV)} &
                                       \multicolumn{3}{|c|}{${\cal H}^{0,12}_{7,6}$, (MeV)} &
                                       \multicolumn{3}{|c}{${\cal H}^{0,12}_{7,6}$, (MeV)}  \\
2j$_a$&2j$_b$&2j$_c$&2j$_d$ & J & T & A=6 &  A=6 &
                                       A=7 &
    $H_{0}^{6,4}$  &    $H_1^{6,5}$  &  $V_2^{6,6}$ &
    $H_{0}^{7,4}$  &    $H_1^{7,5}$  &  $V_2^{7,6}$ &
    $H_{0}^{4,4}$  &    $H_1^{5,5}$  &  $W_2^{7,6}$     \\
 \hline
\hline
 1 &   1 &   1 &   1 &   0 &   1 & -6.369  & -20.528 & -31.866 &      -54.830&  36.762 & -1.626&  -63.336&   33.614 &  -1.241 & -30.500& 11.014 &  -11.638  \\
 1 &   1 &   3 &   3 &   0 &   1 & -3.818  & -2.823 &  -3.104 &             &         & -2.823&         &          &  -3.104 &        &        &   -3.104  \\
 3 &   3 &   3 &   3 &   0 &   1 &  -9.069 & -27.147 & -41.661 &      -54.830&  28.997 & -0.161&  -63.336&   22.555 &   0.401 & -30.500&  6.535 &  -16.728  \\
 1 &   1 &   1 &   1 &   1 &   0 &  -7.526 & -22.822 & -35.152 &      -54.830&  36.762 & -3.921&  -63.336&   33.614 &  -4.526 & -30.500& 11.014 &  -14.923  \\
 1 &   1 &   1 &   3 &   1 &   0 & -1.264  &-0.645 &  -1.025 &  &&                     -0.645&         &          &  -1.025 &        &        &   -1.025 \\
 1 &   1 &   3 &   3 &   1 &   0 &  1.724  & 2.012 &   2.107 &  &&                      2.012&         &          &   2.107 &        &        &    2.107 \\
 1 &   3 &   1 &   3 &   1 &   0 &  -11.183& -27.828 & -41.079 &      -54.830&  32.879 & -4.884&  -63.336&   28.085 &  -4.735 & -30.500&  8.774 &  -18.498 \\
 1 &   3 &   3 &   3 &   1 &   0 & -4.037   & -4.211 &  -4.977 &  &&                     -4.211&         &          &  -4.977 &        &        &   -4.977  \\
 3 &   3 &   3 &   3 &   1 &   0 &  -7.180 & -26.884 & -41.615 &      -54.830&  28.997 &  0.102&  -63.336&   22.555 &   0.448 & -30.500&  6.535 &  -16.681  \\
 1 &   3 &   1 &   3 &   1 &   1 &  -6.239 & -21.419 & -33.875 &      -54.830&  32.879 &  1.524&  -63.336&   28.085 &   2.469 & -30.500&  8.774 &  -11.294  \\
 1 &   3 &   1 &   3 &   2 &   0 & -10.847 & -26.844 & -40.884 &      -54.830&  32.879 & -3.900&  -63.336&   28.085 &  -4.540 & -30.500&  8.774 &  -18.303 \\
 1 &   3 &   1 &   3 &   2 &   1 & -8.292  & -22.951 & -35.742 &      -54.830&  32.879 & -0.007&  -63.336&   28.085 &   0.602 & -30.500&  8.774 &  -13.161  \\
 1 &   3 &   3 &   3 &   2 &   1 &  1.594  & 1.395 &   1.787 &  &&                      1.395&         &          &   1.787 &        &        &    1.787  \\
 3 &   3 &   3 &   3 &   2 &   1 &  -7.165 & -24.892 & -39.188 &      -54.830&  28.997 &  2.094&  -63.336&   22.555 &   2.875 & -30.500&  6.535 &  -14.245  \\
 3 &   3 &   3 &   3 &   3 &   0 &  -9.730 & -29.167 & -44.520 &      -54.830&  28.997 & -2.181&  -63.336&   22.555 &  -2.457 & -30.500&  6.535 &  -19.586  \\
\hline
\end{tabular}
\end{table*}
In Table I we also list the values of $H^{N_{\rm max}=12, \Omega}_{6,2}$ with $\Omega=14$ MeV, so that one can observe how much these values change when
 the correlations up to 6-bodies are included, so as to obtain the values of ${\cal H}^{0,12}_{6,6}$.

 The results presented in Table \ref{V6dec} indicate that the largest parts of the effective Hamiltonian are attributed to
 the interaction among core nucleons (k=0) and the interaction of valence nucleons with the core nucleons (k=1).
 However, these two largest contributions partially cancel each other. The pure two-body part corresponding to
 the interaction of valence nucleons is considerably smaller than the individual core and one-body parts.
 Note that one may re-partition the core and single particle energies by shifting a constant amount from
 $H_0^{A,5}$ to $H_0^{A,4}$. A shift of $\approx 24$ MeV ($\approx 32$ MeV) for A=6 (7) produces core and valence energies where the core matches the $^4$He as in the NCSM with A=4.

\begin{figure}[!ht]
 \includegraphics[scale = 0.20]{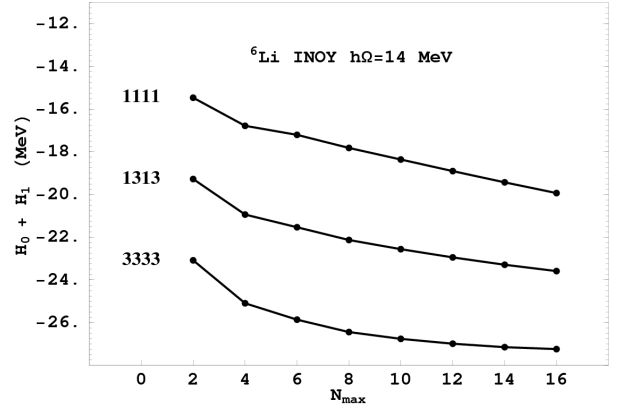}%
 \caption{\label{6li_H_C} The diagonal TBMEs of the sum for the core and one-body parts,
  $\langle ab |H_{0}^{6,4} + H_1^{6,5} | ab \rangle$, for the effective Hamiltonian,
${\cal H}^{0,N_{\rm max}}_{6,6}$, for $^6$Li as a function of $N_{\rm max}$. The corresponding 
curves are labeled by quantum numbers $2j_a2j_b2j_a2j_b$. }
 \end{figure}

To investigate the balance of the pure two-body, $V_2^{6,6}$, core, $H_0^{6,4}$, and one-body, $H_1^{6,5}$,
parts of the effective Hamiltonian with the increase of the size of the original many-body space,
we have plotted the sum of core and one-body parts, $ H_{0}^{6,4}+H_{1}^{6,5}$,
as a function of $N_{\rm max}$ in Fig.\ref{6li_H_C}.
The results in Fig.\ref{6li_H_C} reveal a weak dependence of the sum of the
core and one-body parts of the effective Hamiltonian on $N_{\rm max}$ starting
at $N_{\rm max}=6$. This means that the converged results for core plus one-body parts
of the effective Hamiltonian are closely approached.  The gaps in the curves are governed by the size
 of the spin-orbit splitting  $\epsilon_1-\epsilon_3$.

\begin{figure}[!ht]
 \includegraphics[scale = 0.22]{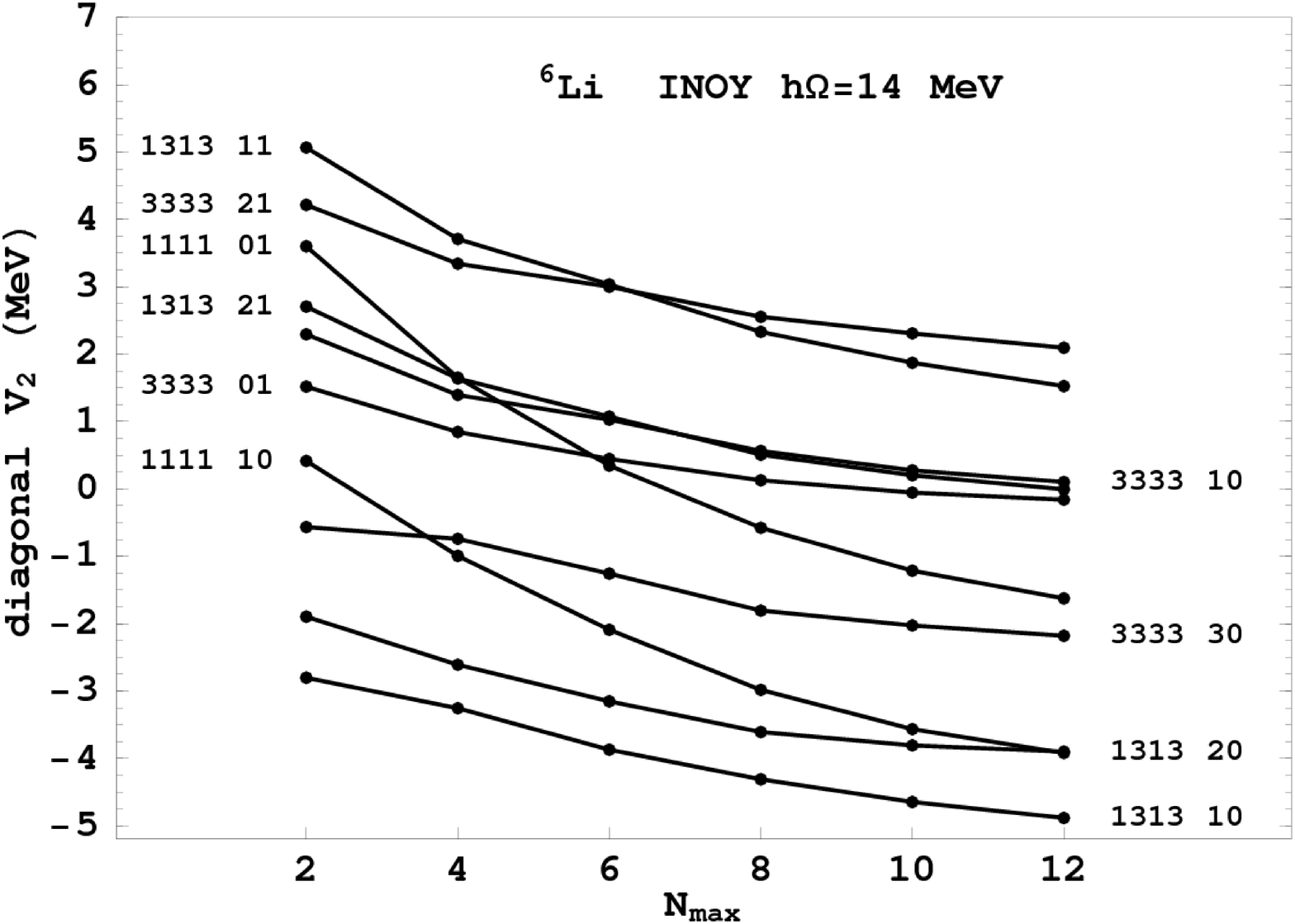}%
 \caption{\label{6li_H_C_SPE} The diagonal pn TBMEs of the  two-body part,
 $\langle ab |V_{2}^{6,6}| ab \rangle_{JT}$,  of the effective Hamiltonian,
 ${\cal H}^{0,N_{\rm max}}_{6,6}$, as a function of $N_{\rm max}$. The corresponding
curves are labeled by quantum numbers $2j_a2j_b2j_a2j_b$ and JT.}
 \end{figure}

\begin{figure}[!ht]
 \includegraphics[scale = 0.20]{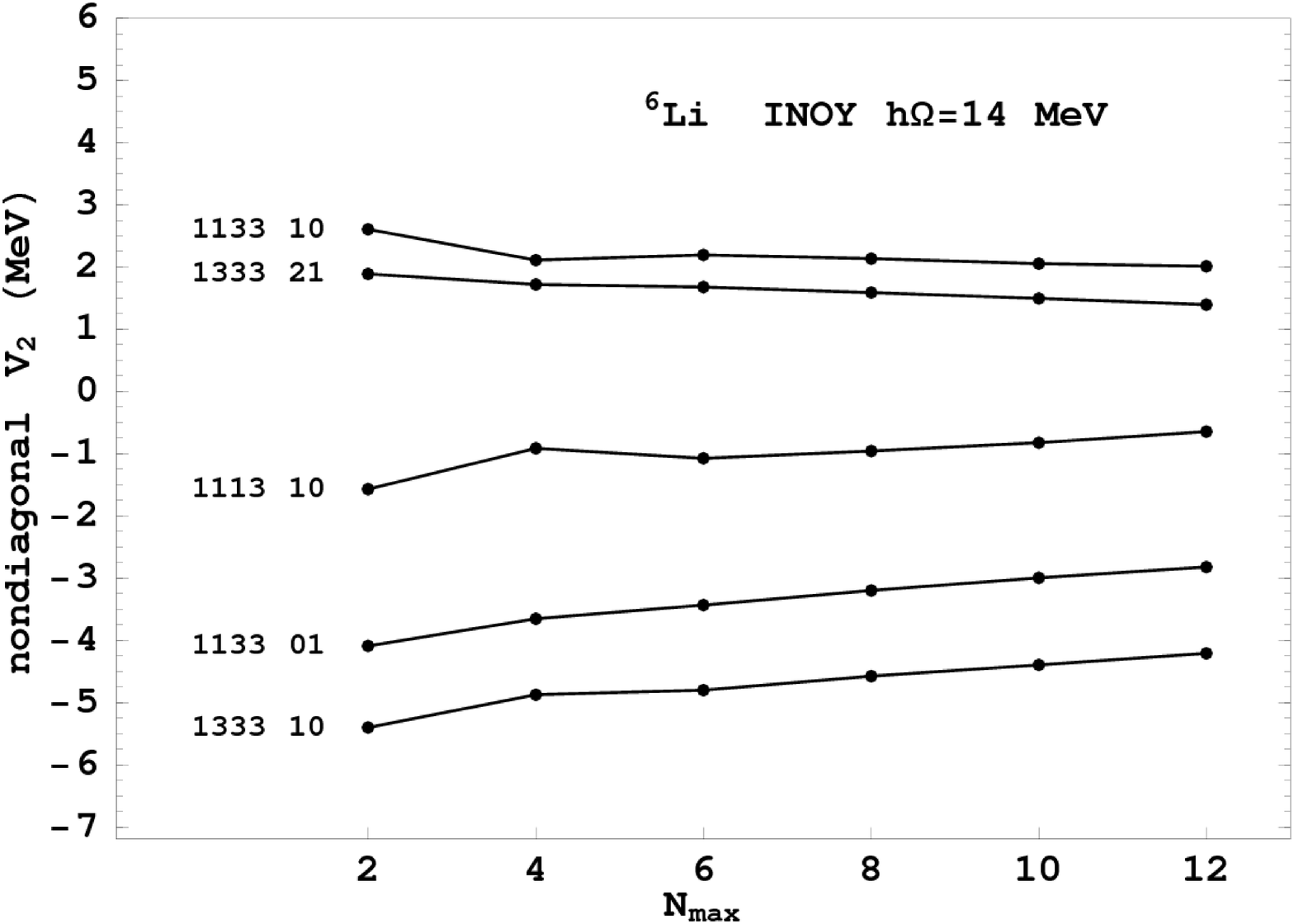}%
 \caption{\label{6li_H_non} The non-diagonal pn TBMEs of the  two-body part,
 $\langle ab |V_{2}^{6,6}| cd \rangle_{JT}$,  for the effective Hamiltonian,
${\cal H}^{0,N_{\rm max}}_{6,6}$, as a function of $N_{\rm max}$. The corresponding
curves are labeled by quantum numbers $2j_a2j_b2j_c2j_d$ and JT.}
 \end{figure}

 Plotting the diagonal pn TBMEs of the residual two-body part, $V_2^{6,6}$, of the effective Hamiltonian in
Fig.\ref{6li_H_C_SPE}, we observe, that they exhibit stronger dependence than the
core plus one-body parts with increase of $N_{\rm max}$.
From Fig.\ref{6li_H_C_SPE} we observe that the T=0 TBMEs are, on average, attractive, while the T=1 TBMEs are
repulsive.  Starting at $N_{\rm max}$=6 the two-body part shows smooth regularity. The results for nondiagonal
matrix elements, shown in Fig. \ref{6li_H_non}, indicates smooth, regular changes towards smaller absolute
values of these TBMEs. We note that slow convergence of TBMEs with increasing $N_{\rm max}$ reminds us of
earlier treatment of core polarization \cite{Bar70,Var73}, where we observe slow convergence with "improved" treatments of core-polarization within perturbation theory.

\subsection{ Two-body valence cluster approximation for $A>6$}

The VCE given by the Eq.(\ref{hexp}) would require a three-body part
V$_3^{7,7}$ of the p-shell effective interaction ${\cal H}^{0,N_{\rm max}}_{7,7}$ to reproduce
exactly the NCSM results for A=7 nuclei:
\begin{equation}
\label{hncsm7}
{\cal H}^{0,N_{\rm max}}_{A=7,a_1=7} = H_{0}^{7,4} + H_{1}^{7,5} + V_{2}^{7,6} + V_{3}^{7,7}.
\end{equation}
Therefore, it is worth knowing how good the 2BVC approximation
for A=7 as well as for $A>7$ is. To test the  2BVC approximation, we have constructed
the ${\cal H}^{0,N_{\rm max}}_{A=7,a_1=6}$ Hamiltonian, using  Eq.(\ref{heff2}), and expanded it in
terms of zero-, one- and two-body valence clusters, {\em i.e.} omitting the three-body part:
\begin{equation}
\label{hdec1}
{\cal H}^{0,N_{\rm max}}_{A=7,a_1=6} = H_{0}^{7,4} + H_{1}^{7,5} + V_{2}^{7,6}.
\end{equation}
 In other words, we have first performed NCSM calculations for the $a_1$-body  systems ($a_1=4,5,6$)
with the $H^{N_{\rm max},\Omega}_{A=7,2}$ Hamiltonian. Thus, $H_{0}^{7,a_1=4}$ is the $^4$He ``core'' energy and
   $H_{1}^{7,a_1=5}$ is the one-body part determined as in Eqs.(\ref{v1})-(\ref{v1spe}), but with A=7; and
$V_{2}^{7,a_1=6}$ is obtained by subtracting $H_{0}^{7,4} + H_{1}^{7,5}$
from ${\cal H}^{0,N_{\rm max}}_{A=7,a_1=6}$.

\begin{figure}[!ht]
 \includegraphics[scale = 0.20]{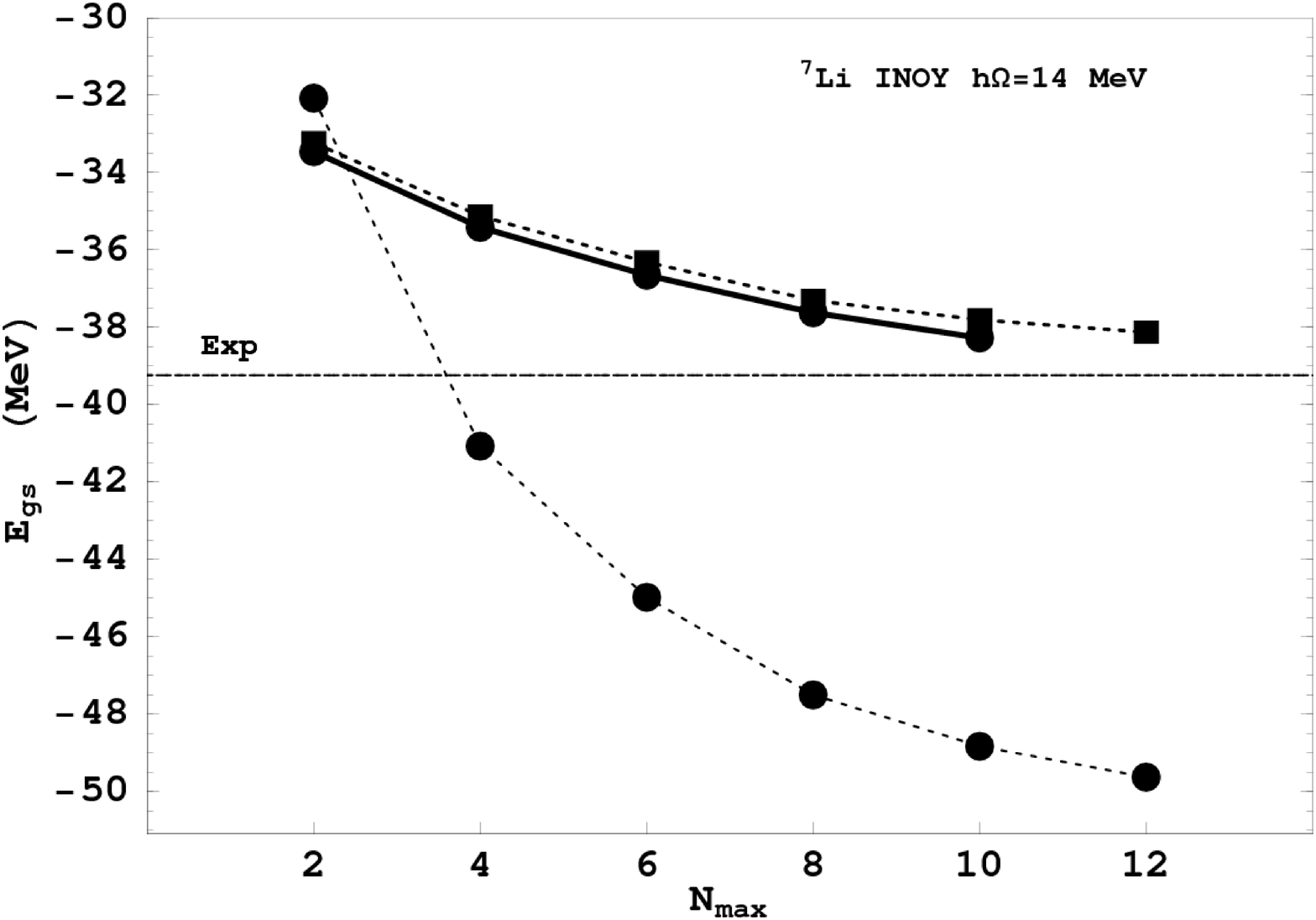}%
 \caption{\label{7li_bind} The ground state energy, $E_{\rm gs}$,  of $^7$Li as a function of $N_{\rm max}$. The NCSM results
with the $H^{N_{\rm max},\Omega}_{A=7,2}$ Hamiltonian  are shown by filled circles connected with the
solid line.
  The SSM results with the effective ${\cal H}^{0,N_{\rm max}}_{7,6}$ Hamiltonian decomposed
according to Eq.(\ref{hdec1}) are shown by squares connected with the dashed line.  The SSM results with the
effective ${\cal H}^{0,N_{\rm max}}_{7,6}$ Hamiltonian decomposed
according to Eq.(\ref{hdec2}) are shown by filled circles connected with a dashed line. }
 \end{figure}

 The resulting parts of the ${\cal H}^{0,N_{\rm max}}_{A=7,6}$ Hamiltonian are given in
Table \ref{V6dec}. Comparing the TBMEs for A=6 and A=7 (Table \ref{V6dec}), we find that they differ considerably.
 There is a big change separately for the core and one-body parts, but weaker changes for the
two-body parts, which tend to become larger in magnitude with increasing A.  We have then performed SSM calculations for the ground state energy of  $^7$Li,
using the zero-, one- and
two-body parts in Eq.(\ref{hdec1}). Namely, the one- and two-body parts were employed in a SSM calculation of the
 ground and excited states energies of the valence nucleons in the p-shell, {\em i.e.}, $0\hbar\Omega$ space, to which
 the $^4$He core energy, $H_{0}^{7,4}$, was added, in order to yield the total energies. These calculations were
 repeated for $N_{\rm max}=0,2,...10$. Next we carried out NCSM calculations for $^7$Li with
 $H^{N_{\rm max},\Omega}_{A=7,2}$ for the same values of $N_{\rm max}$. The SSM and NCSM results for the
 ground-state energy are shown in Fig.\ref{7li_bind}.

\begin{figure}[!ht]
\includegraphics[scale = 0.20]{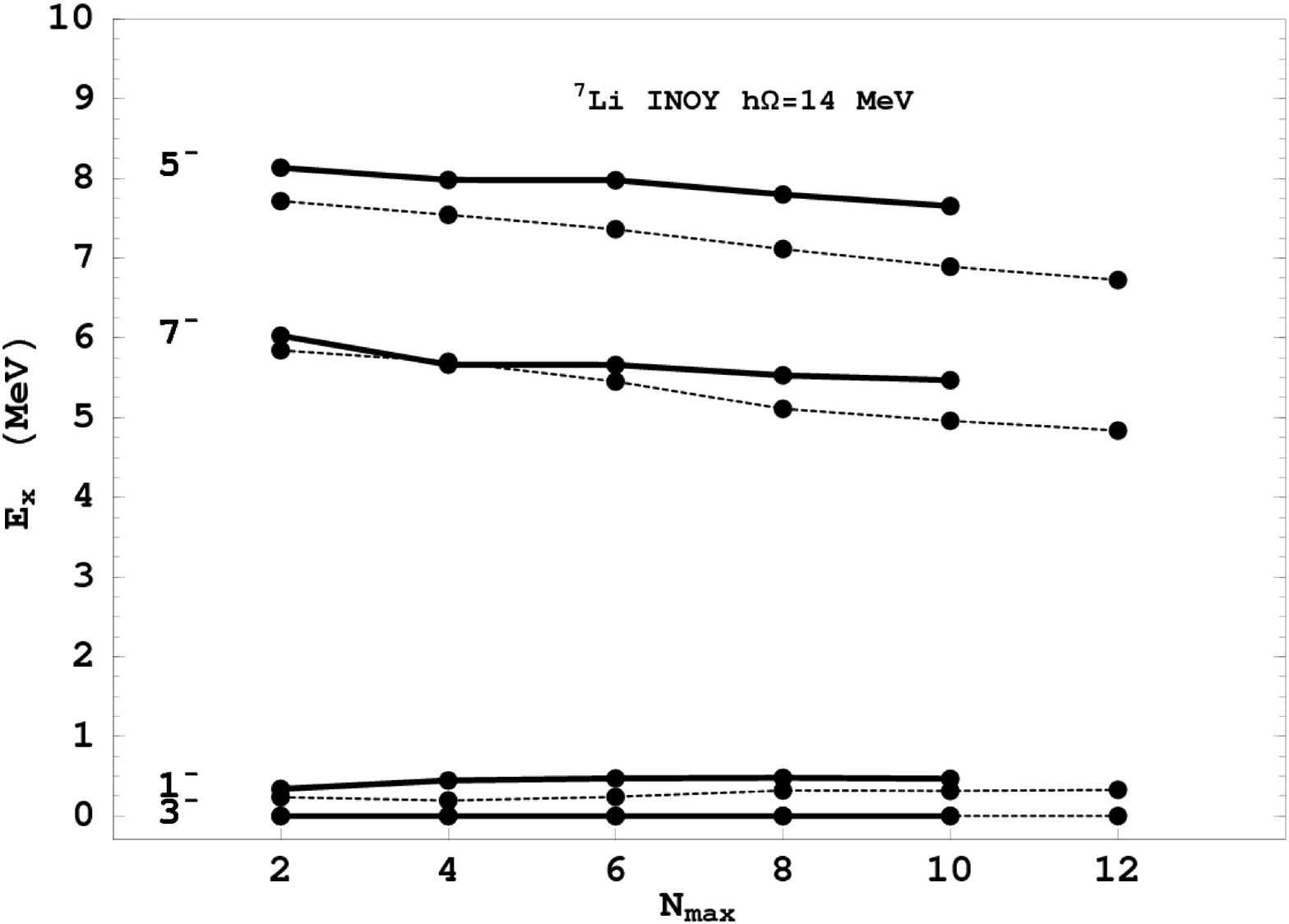}%
\caption{\label{7li_spectra} NCSM (solid line) and SSM (using Eq.(\ref{hdec1}), dashed line) spectra for $^7$Li.
The states with spin J are marked
by 2J. }
 \end{figure}

It is also of interest  to find out what would be the result if we
take the fixed core and one-body parts at values which are appropriate for
 the $a_1=4$ and $a_1=5$ systems, respectively, because this is analogous to what is done in the SSM to determine
energies relative to an inert core.  To do this we adopt an alternative two-body VCE,
 which assumes that the core and one-body parts are A independent, {\em i.e.},
\begin{equation}
\label{hdec2}
{\cal H}^{0,N_{\rm max}}_{A,6} = H_{0}^{4,4} + H_1^{5,5} + W_2^{A,6},
\end{equation}
similar to the SSM convention given by Eq.(\ref{hssm}). We have then performed another set of SSM calculations
for A=7 in the same manner as described previously, but using the decomposition given in Eq.(\ref{hdec2}). To distinguish between the two-body part of the VCE given
by the Eqs.(\ref{hexp}) and (\ref{hdec2}), we have introduced the new notation, $W_2^{A,6}$,  in Eq.(\ref{hdec2}).
The Hamiltonian ${\cal H}^{0,12}_{7,6}$ expanded according to the
Eq.({\ref{hdec2}}) is shown in last three columns of Table \ref{V6dec} and the corresponding results are depicted in
Fig.\ref{7li_bind} by the dots connected with a dashed line.
Figure \ref{7li_bind} indicates that for light systems a realistic balance of core, one-body and two-body parts of
the effective interaction may be achieved only when both the core and one-body parts are mass-dependent, contrary to
earlier approaches.  A-independent
core and one-body parts lead to a very strong two-body part for the valence nucleons and, subsequently, to
drastic overbinding.
It is obvious, that, in order to compensate for such an effect one would need to introduce a strongly repulsive three-body
effective interaction with an unrealistic strength of about 10 MeV. Although,
the effect on the spectrum is smaller, the VCE with the A-dependent core and one-body parts also yields
 better agreement with the exact NCSM results for the excited states. The corresponding low-energy spectrum
of $^7$Li obtained with the NCSM and the A-dependent SSM (using the values in columns 12,13 and 14 of
Table \ref{V6dec}) are compared in
Fig.\ref{7li_spectra}.
The differences observed in Figs.\ref{7li_bind} and \ref{7li_spectra} for the ground state and excited states,
respectively, may be attributed to the neglected three-body part of the effective interaction  at the two-body
valence cluster level.

\begin{figure}[!ht]
\includegraphics[scale = 0.20]{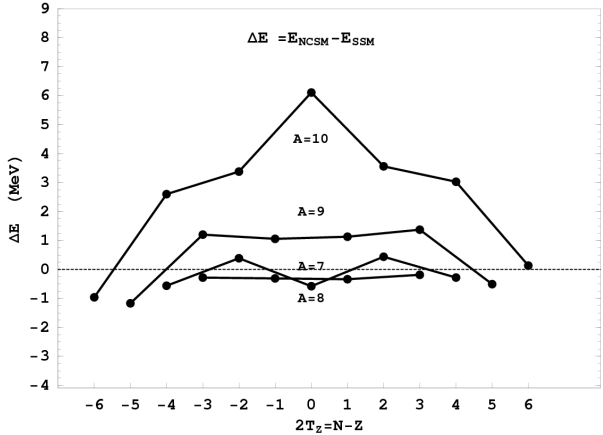}%
\caption{\label{a7-10_bind} The difference of the NCSM and SSM  (Eq.(\ref{hdecg}) ground state energies for
 different values of mass
number A as a function of isospin projection $T_z=(N-Z)/2$.}
 \end{figure}

 We have generalized the 2BVC expansion procedure of Eq.(\ref{hdec1}) for arbitrary mass number A,
\begin{equation}
\label{hdecg}
{\cal H}^{0,N_{\rm max}}_{A,a_1=6} = H_{0}^{A,4} + H_{1}^{A,5} + V_{2}^{A,6},
\end{equation}
 and applied it to the A=7,8,9, and 10 isobars for $N_{\rm max}=6$.
 The difference of the NCSM and SSM  ground state energies for different mass number
A is plotted as a function of isospin projection $T_z=(N-Z)/2$ in Fig.\ref{a7-10_bind}.
Figure \ref{a7-10_bind} shows that the three-body and higher-body correlations become more important with
increasing mass number. There is also a very strong isospin dependence of the obtained results. For the highest isospin values
the SSM systematically underbinds nuclei in comparison to the NCSM and higher-body correlations appear to be small for systems
 containing only valence neutrons. However, there is an opposite effect in the vicinity
of the $N=Z$ line where SSM yields considerably more binding energy than the NCSM.

 Thus, the residual  $a_1$-body correlations with $a_{\rm 1} \ge 3$ in the p-shell play an important role for $A \ge 7$
nuclei in terms of total binding energy.

\subsection{Beyond the two-body valence cluster expansion}

The analysis of the A=7 systems may allow us to derive an effective three-body Hamiltonian for the p-shell and to
 give an idea
about the strength of the three-body interaction. To derive the three-body effective Hamiltonian,
we employ the three-body valence cluster expansion (3BVC) approximation,
\begin{equation}
\label{hvdec3}
{\cal H}^{0,N_{\rm max}}_{A,a_1=7} = H_{0}^{A,4} + H_{1}^{A,5} + V_{2}^{A,6} + V_{3}^{A,7},
\end{equation}
which is the exact one for $A=7$ systems. Comparing Eqs.(\ref{hdec1}) and (\ref{hvdec3}), we find the the following
result for the three-body part $V_{3}^{A,7}$  of the effective Hamiltonian:
\begin{equation}
\label{v3}
V_3^{A,7} = {\cal H}^{0,N_{\rm max}}_{A,7} - {\cal H}^{0,N_{\rm max}}_{A,6}.
\end{equation}
Using Eq.(\ref{heff2}), we derive the A=7 Hamiltonian, ${\cal H}^{0,N_{\rm max}}_{7,7}$,
employing $a_1=7$ NCSM eigenvectors and eigenvalues, obtained with the $H^{N_{\rm max},\Omega}_{7,2}$
 interaction. The same procedure is then repeated to calculate the A=7 Hamiltonian, ${\cal H}^{0,N_{\rm max}}_{7,6}$,
employing $a_1=6$ NCSM eigenvectors and eigenvalues, obtained with the $H^{N_{\rm max},\Omega}_{6,2}$
 interaction.   Then, the residual three-body part $V_3^{7,7}$ is calculated according to Eq.(\ref{v3}).
The same scheme can be applied for $A>7$ systems taking appropriate values of A in Eq.(\ref{v3}).

As an example, the neutron (nnn) T=3/2 matrix elements of the resulting three-body effective p-shell Hamiltonian for A=7 and
$N_{\rm max}=6$  are given in Table \ref{V3bodyt3}.
 \begin{table}[tbp]
 \caption{\label{V3bodyt3} The 3-body T=3/2 parts of the p-shell effective Hamiltonian, ${\cal H}^{0,N_{\rm max}}_{7,7}$,
obtained from an $N_{\rm max}=6$ NCSM calculation for  $^7$He is shown in column 9. The 3-body nnn parts of the p-shell effective Hamiltonians, ${\cal H}^{0,N_{\rm max}}_{A,7}$, for
A=8, 9 and 10 are shown in columns 10,11 and 12, respectively.}
 \begin{tabular}{ccc|ccc|cc|cccc}
 \hline\hline
&&&&&&&&\multicolumn{4}{|c}{${ V}^{3}_{A,7}$, (MeV)} \\ \cline{9-12}
2j$_a$&2j$_b$&2j$_c$&2j$_d$& 2j$_e$&2j$_f$  & 2J & 2T &                      $A=7$ &
                                                                            $A=8$  &
                                                                            $A=9$  &
                                                                            $A=10$ \\
   &     &     &     &     &     &  &  &  nnn  & nnn  & nnn & nnn \\
 \hline
 3 &   3 &   1 &       3 &   3 &  1 & 1  &   3 &   -0.055  &  0.181 &  0.354 &  0.471\\
 3 &   3 &   3 &       3 &   3 &  3 & 3  &   3 &   -0.366  & -0.181 & -0.080 & -0.026 \\
 3 &   3 &   1 &       3 &   3 &  1 & 3  &   3 &   -0.504  & -0.280 & -0.126 & -0.030 \\
 3 &   1 &   1 &       3 &   1 &  1 & 3  &   3 &   -0.306  & -0.197 & -0.081 &  0.010 \\
\hline
3 &   3 &   3 &       3 &   3 &  1 & 3  &   3 &     0.290  &   0.281  &  0.270 &  0.261 \\
3 &   3 &   3 &       3 &   1 &  1 & 3  &   3 &    -0.246  &  -0.202  & -0.165 & -0.135\\
3 &   3 &   1 &       3 &   1 &  1 & 3  &   3 &     0.388  &   0.356  &  0.317 &  0.283 \\
\hline
3 &   3 &   1 &       3 &   3 &  1 & 5  &   3 &    -0.209  & -0.038 & 0.066 & 0.124 \\
\hline
\end{tabular}
\end{table}
On average, the nnn $T=3/2$ Three-Body Matrix Elements (3BMEs) are attractive for A=7.
They are approximately an order of magnitude smaller in absolute value than the related $T=1$ TBMEs for A=7
(see Table \ref{V6dec}) and have an opposite sign. Performing the same procedure, we have obtained the 3BMEs for the
A=8, 9 and 10 systems, which are also listed in Table \ref{V3bodyt3}. Comparing nnn 3BMEs for different A, we note that
diagonal 3BMEs become more repulsive, while there are only small changes for non-diagonal 3BMEs; however their magnitudes become smaller for larger mass. This is in contrast to what we observed in the previous section for the two-body effective interaction.

\begin{figure}[!ht]
\includegraphics[scale = 0.20]{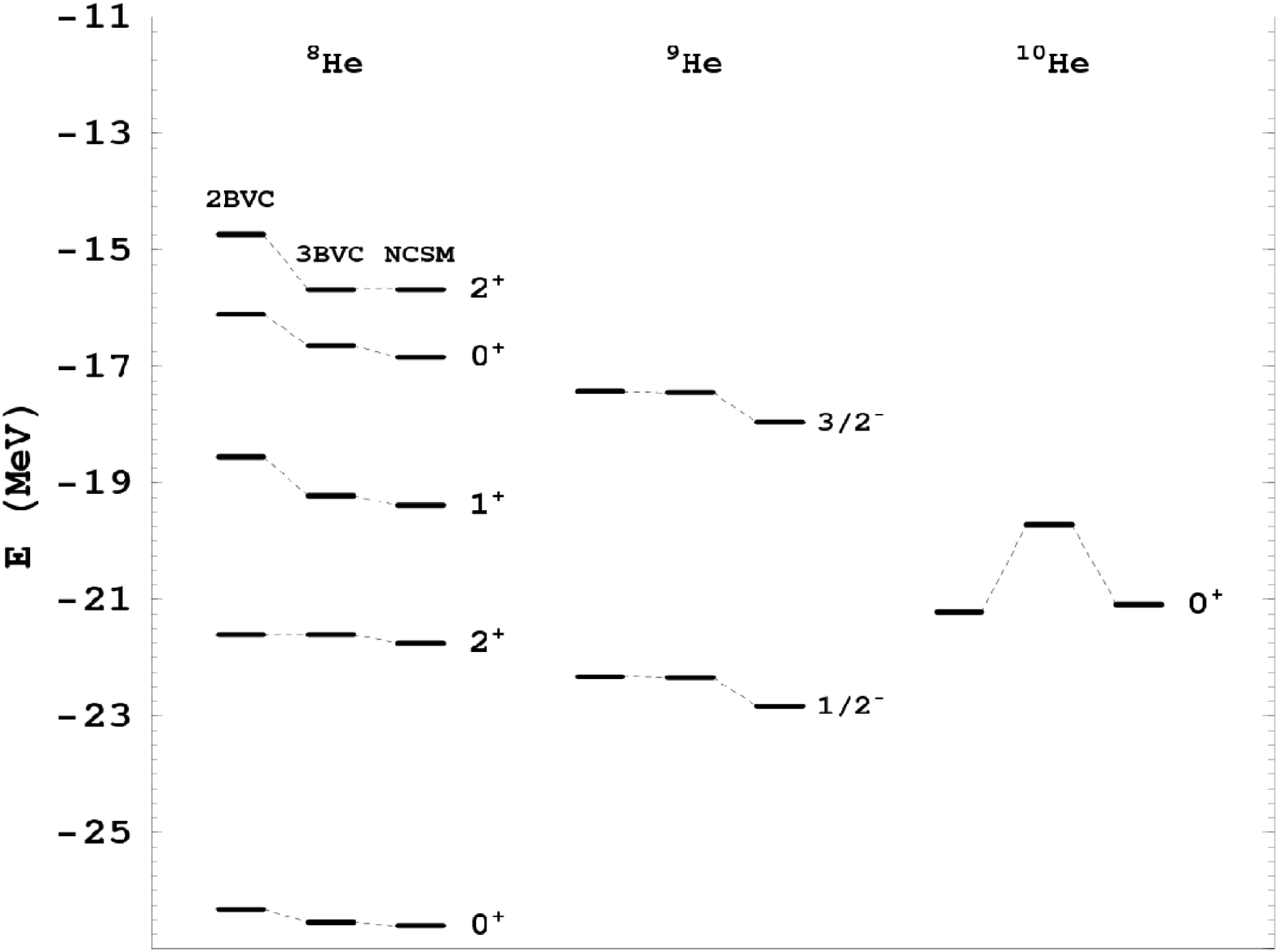}%
\caption{\label{he8-9} Comparison of spectra for $^8$He, $^9$He and $^{10}$He from
 SSM calculations using the effective 2BVC and 3BVC Hamiltonians and from exact NCSM calculation for $N_{\rm max}=6$ and $\Omega$=14 MeV using
 the INOY interaction.}
 \end{figure}

The T=3/2 3BMEs can be represented in terms of T=1 TBMEs using the coefficients of fractional parentage (CFP)
for the 3-body to 2-body reduction problem. Following this idea, we have calculated 3-body corrections for the corresponding TBMEs
using T=3/2 3BMEs shown in Table II. It is worth noting, that this is not an exact way to treat the 3-body degrees of freedom
but an approximation which estimates average 3-body effect.  Using the 3-body corrected neutron TBMEs, we have performed SSM calculations
for $^8$He, $^9$He and  $^{10}$He, which have no valence protons and 4, 5 and 6 valence neutrons, respectively, in the p-shell.
Since there are only valence neutrons in the case of He isotopes, only the T=3/2 three-body coupling is possible,
and, thus, the T=1/2 3BMEs are not required for calculations.
 As an example, the results of the SSM calculations for $^8$He, $^9$He and  $^{10}$He with effective
interactions obtained in 2BVC and 3BVC approximations from INOY interaction are compared to exact NCSM results in Table \ref{he8}
and Fig. \ref{he8-9}.

 \begin{table}[tbp]
 \caption{\label{he8} Results for $^8$He, $^9$He and $^{10}$He from SSM calculations with the
 effective 2BVC and 3BVC Hamiltonians and from exact NCSM calculation for $N_{\rm max}=6$
 with the INOY interaction.}
 \begin{tabular}{c|ccc|c|ccc}
 \hline \hline
 $J_i^\pi$ &  \multicolumn{3}{c|}{E($^8$He), (MeV)} & $J_i^\pi$ &  \multicolumn{3}{|c}{E($^9$He), (MeV)}  \\
\hline
     &  2BVC     & 3BVC         &   NCSM &        &  2BVC     & 3BVC  &   NCSM   \\
\hline
 $0_1^+$  &  -26.323    & -26.542       & -26.604  & $1/2_1^-$&  -22.328       & -22.342    & -22.835        \\
 $2_1^+$  &  -21.608    & -21.609       & -21.752  & $3/2_1^-$&  -17.429       & -17.452    & -17.961 \\
 \cline{6-8}
 $1_1^+$  &  -18.555    & -19.224       & -19.386  &          & \multicolumn{3}{|c}{E($^{10}$He), (MeV)} \\
 \cline{6-8}
 $0_2^+$  &  -16.108     & -16.644       & -16.843  & $0^+$    & -21.219 & -19.720 & -21.086   \\
 $2_2^+$  &  -14.736     & -15.681        & -15.682   &&&&\\
\hline
\end{tabular}
\end{table}

\begin{figure}[!ht]
\includegraphics[scale = 0.20]{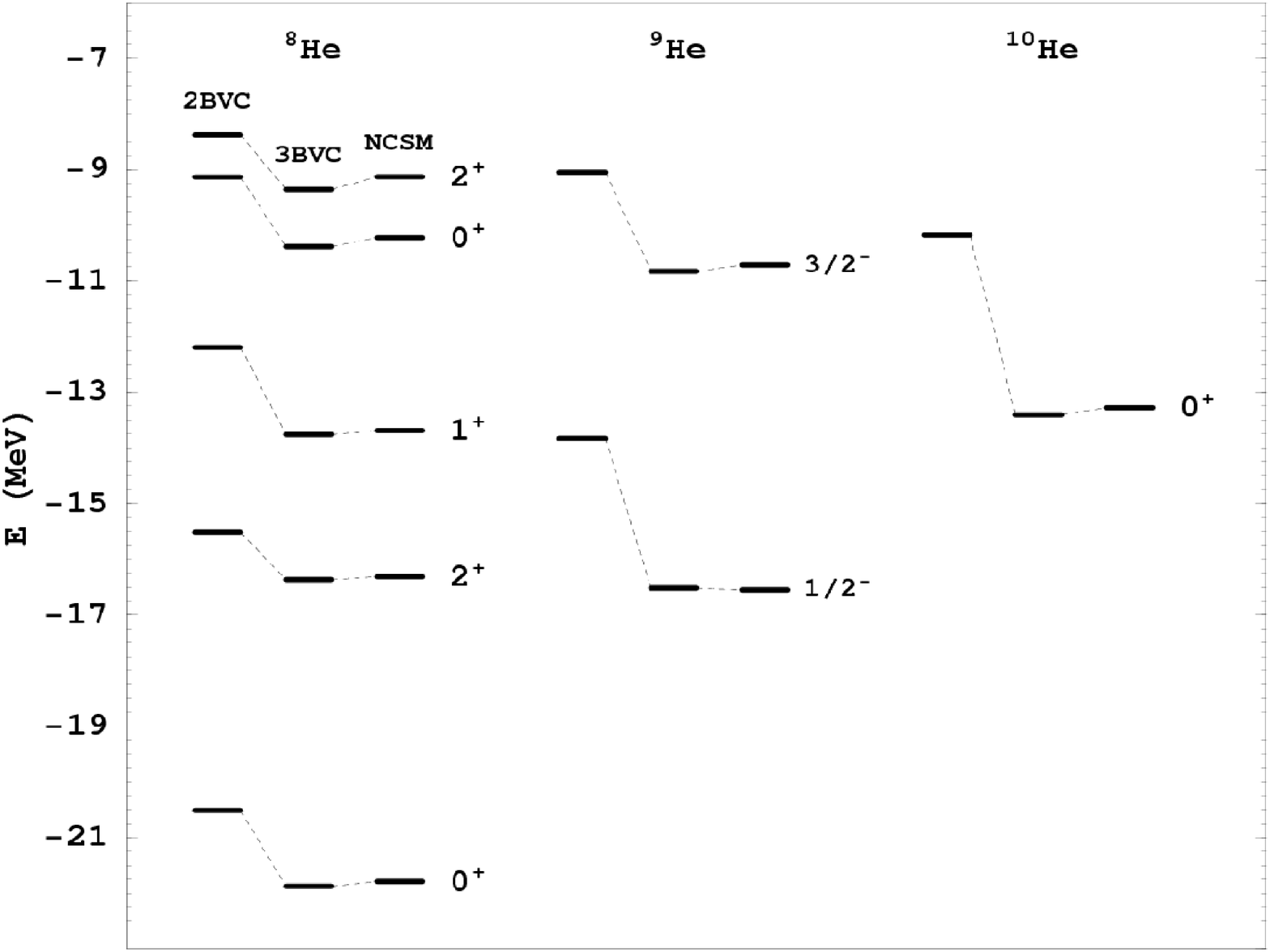}%
\caption{\label{he8-9cd} Comparison of spectra for $^8$He, $^9$He and $^{10}$He from
 SSM calculations using the effective 2BVC and 3BVC Hamiltonians and from exact NCSM calculation for $N_{\rm max}=6$ and $\Omega$=20 MeV using
 CD-Bonn interaction.}
 \end{figure}

Obtained results indicate that accounting for the effective 3-body interactions considerably improves the agreement with the exact
NCSM for the $^8$He, does not bring much change for $^9$He and yields worse results for $^{10}$He (see Fig.\ref{he8-9}).
Performing a similar calculation with the effective interaction obtained in the 3BVC approximation starting from the CD-Bonn
interaction \cite{Mac96}, we obtained results which are shown in Fig.\ref{he8-9cd}.
Note, that the effective CD-Bonn interaction constructed in the 2BVC approximation considerably underbinds the He isotopes in comparison to the
 exact NCSM results. The subsequent employment of the 3BVC approximation compensates these large differences and yields much better results for
 $^{10}$He. However, to draw more quantitative conclusion about the 3-body and higher-body
effective interactions, one needs to perform exact diagonalization using the 3BMEs. We will evaluate this effect in future studies.

\section{Conclusion}

Within the NCSM approach we can calculate, by exact projection, full A-nucleon dependent TBMEs (and 3BMEs).
These  A-dependent  TBMEs (and 3BMEs) can be separated into core, one-body and two-body (and three-body) parts, all of which are also A-dependent, contrary to the SSM approach. When these A-dependent effective one- and
two-body (and three-body) interactions are employed in SSM calculations, they exactly reproduce full NCSM calculations for A=6 (A=7) isobars
and yield results in good agreement with full NCSM calculations for $A>7$ performed in large basis spaces. Our results for $A>7$, which include the 3-body effective interaction, indicate that 3- and higher-body effective interactions may play an important role in determining their binding energies and spectra. Future
investigations will be extended to include effective 3-body interactions exactly and to explore other physical operators, such as transition operators and EM moments.

\section{Acknowledgments}
We thank the Institute for Nuclear Theory at the University of
    Washington for its hospitality and the Department of Energy
    for partial support during the development of this work.
B.R.B. and A.F.L. acknowledge partial support of this
work from NSF grants PHY0244389 and PHY0555396; P.N.
acknowledges support in part by the U.S. DOE/SC/NP
(Work Proposal N. SCW0498) and U.S Department of Energy Grant DE-FG02-87ER40371;
J.P.V. acknowledges support from U.S. Department of Energy
Grants DE-FG02-87ER40371 and DE-FC02-07ER41457; and
the work of I.S. was performed under the auspices
of the U.S. DOE.  Prepared by
LLNL under Contract DE-AC52-07NA27344. B.R.B. thanks the Gesellschaft
f\"ur Schwerionenforschung mbh Darmstadt, Germany,
for its hospitality during the preparation of this
manuscript and the Alexander von Humboldt Stiftung
for its support.

\end{document}